\documentclass[showpacs,showkeys,prbleq,12pt]{revtex4}
\usepackage{graphicx}
\usepackage{dcolumn}
\usepackage{bm}
\usepackage{amsmath}
\usepackage{graphicx}
\usepackage{paralist}

\begin{document}

\title{Study of the first-order phase transition in the classical and quantum random field Heisenberg model on a simple cubic lattice}
\author{J. Ricardo de Sousa$^{a,b}$, Douglas F. de Albuquerque$^{c}$, Alberto S. Arruda$^{d}$}
\affiliation{$^{a}$Departamento de F\'{\i}sica, Universidade Federal do Amazonas, 69077-000, Manaus-AM, Brazil\\
$^{b}$National Institute of Science and Technology for Complex Systems, 3000, Japiim, 69077-000, Manaus-AM, Brazil\\
$^{c}$Departamento de Matem\'{a}tica, Universidade Federal de Sergipe, 49100-000, S\~{a}o Cristov\~{a}o-SE, Brazil\\
$^{d}$Instituto de F\'{\i}sica, Universidade Federal de Mato Grosso, 78060-900, Cuiab\'{a}-MT, Brazil}

\begin{abstract}
The phase diagram of the Heisenberg ferromagnetic model in the presence of a magnetic random field (we have used bimodal distribution) of spin $S=1/2$ (quantum case) and $S=\infty $ (classical case) on a simple cubic lattice is studied within the framework of the effective-field theory in finite cluster (we have chosen $N=2$ spins). Integrating out the part of order parameter (equation of state), we obtained an effective Landau expansion  for the free energy written in terms of the order parameter $\Psi (m)$. Using Maxwell construction we have obtained the phase diagram in the $T-H$ plane for all interval of field.  The first-order transition temperature is calculated by the discontinuity of the magnetization at $T_{c}^{\ast }(H)$, on the other hand in the continuous transition the magnetization is null at $T=T_{c}(H)$. At null temperature ($T=0$) we have found the \textbf{coexistence} field $H_{c}=3.23J$ that is independent of spin value. The transition temperature $T_{c}(H)$ for the classical case ($S=\infty $), in the $T-H$ plane, is larger than the quantum case ($S=1/2$).
\end{abstract}

\maketitle

\section{Introduction}

Phase transitions are one of most interesting phenomena that occurs in nature. Many systems have phase transitions in critical regions and it is widely known that the classic Ising model (and others) displays a second order temperature driven phase transition. In particular, phase transitions and the critical behaviors of the random field Ising model (RFIM) were studied extensively in the last years, see \cite{Belanger1998, F.2002, Belanger2000} and references therein. The RFIM leads to a number of challenging problems in the physics of disordered systems\cite{  F.2002, Belanger2000, Shelton2004}. There are two basic types of disorder in spin models: \begin{inparaenum}[i)] \item disordered bonds (spin-glass models) and \item site disorder (randomness of the applied magnetic field in the RFIM).\end{inparaenum} Mean-field theory has been one of several techniques used to study the RFIM. Although the mean-field version of the RFIM is much easier, there are some open questions about the behavior of the RFIM with more realistic, short-rgange interactions, which still motivate experimental and theoretical investigations\cite{F.2002, Belanger2000}. The lower critical dimension and the existence of an ordered phase in the three-dimensional case, have been rigorously established by mathematical proofs\cite{Imbrie1984, prl351975I}. However, the existence of a tricritical point (TCP) for a double-$\delta$ distribution of random fields, in accordance with mean-field results, is still under question( see \cite{mprl231997A, Mattis1985} and references therein).

\hspace{1em}The RFIM is revelant for the description of several physical situations, as example: \begin{inparaenum}[i)]\item for the structural phase transitions in random alloys, \item  for the phase transitions in commensurate charge-density-wave systems with impurity pinning and \item in binary fluid mixtures in random porous media\end{inparaenum}. Random fields have been used to mimic frustration introduced by the disorder in interacting many body systems and for explaining several aspects of electronic transport in disordered insulators\cite{Efros1975} and in systems near the metal-insulator transition \cite{Kirkpatrick1994,Pastor1999}. On the other hand, the physics of hysteresis, of the avalanche behavior, and of the origin of self-organized criticality \cite{Pastor2002}, has been explained by resorting to the analysis of the non-equilibrium behavior of suitable RFIM. There is a new class of problems related to the self-generated glassy behavior, which has been explained instead in terms of a spin model in infinitesimal random fields \cite{Mezard1999}, and more recently, the RFIM has  been
 employed to describe critical behavior of amorphous magnetic
 systems,  such as thin films  and critical surface behavior  of the amorphous
 semi-infinite  systems~\cite{jmmm2182000A,jmmm2192000A}.%

In the last years, a new effective field theory (EFT) has been used to study  second-order phase transition of  both classical and quantum spin models, and tricritical  point in the phase diagram,  which leads to useful qualitative insights for the critical behavior. The EFT method uses the Callen-Suzuki identities\cite{1} as a starting point  and utilizes the differential operator technique, developed by Honmura and Kaneyoshi\cite{2}. It provides a hierarchy of  approximations to obtain thermodynamic properties of magnetic models.  One can continue these series of aproximations  considering increasing  clusters  which leads to better results.  The exact solution would be obtained by considering an infinite cluster. However, by using relatively small clusters that contain the topology of the lattice, one can obtain a reasonable description of thermodynamic properties as it will be shown below.

Several spin models,  such as Blume-Capel~\cite{3}, random field Ising~\cite{4,bou},  Heisenberg~\cite{5, douglas, 6}, Ising metamagnet~\cite{7,zuk} and  Ising with four-spin couplings~\cite{8,kaney} models, have been treated by using EFT. In these works, the first-order line could not be obtained due to the absence of an expression for the free energy. Therefore, only second-order lines and tricritical points were analyzed. In particular, Fittipaldi and Kaneyoshi~\cite{9} have used the EFT approach to study the phase diagram of the Blume-Capel model with spin-$1$ on a  two-dimensional lattice. The position of the first-order transition was obtained from the isotherms in the $m-H$ plane (where $m$ and $H$ are the magnetization and the magnetic field, respectively) applying the Maxwell equal area construction.  The first-order lines obtained in Ref. [28] are not correct, since in the limit $\alpha =-0.50$ (where $\alpha =J^{\prime }/J $, and $J^{\prime }(J)$ is the biquadratic (bilinear) coupling) at $T=0$ the exact value is  $D/J=-0.75$, and the value presented in Ref. [28] was $D/J=-0.50$ (see Fig. 1).

Recently, de Albuquerque, \textit{et al}.\cite{5,douglas} have studied the phase diagram of the random field classical Heisenberg model (RFHM) on a simple cubic lattice. Oubelkacem, \textit{et al}.\cite{6} extended the calculation to treat the quantum spin-$1/2$ random field Heisenberg model and obtained only second-order lines and tricritical points \cite{5, douglas, 6}. The purpose of this work is to discuss the complete phase diagram (entire range of the field) in the $T-H$ plane of the random field Heisenberg model on a simple cubic lattice by using EFT in two-spins cluster (EFT-2).

In the present work, our goal is to propose a functional for free energy, in order to obtain the first-order line in the phase diagram in the $T-H$ plane for the random field classical and quantum spin-$1/2$ Heisenberg model on a simple cubic lattice. The outline of this paper is as follows: the model and formalism are developed in section II, and the results and conclusions are discussed in section III.

\section{Model and Formalism}

In order to obtain the free energy, we developed the calculations to treat the phase diagram of the RFHM with classical ($S=\infty $) and quantum ($S=1/2$) spins. The RFHM is described by the following Hamiltonian:%
\begin{equation}
\mathcal{H}=-J\sum_{\langle {ij}\rangle }\mathbf{S}_{i}\cdot \mathbf{S}%
_{j}-\sum_{i}H_{i}S_{i}^{z},  \tag{1}
\end{equation}%
where the first sum is carried out only over pairs of nearest-neighboring sites with the interaction $J$. Also $S_{i}^{z}$ is the $z$-component of the spin operator (vector) $\mathbf{S}_{i}=(S_{i}^{x},S_{i}^{y},S_{i}^{z})$ at site $i$. For the classical case\cite{stanley} we  consider the normalization condition $\sum\limits_{\mu =x,y,z}(S_{i}^{\mu })^{2}=3$
 and for the quantum case $\mathbf{S}_{i}$ is  now considered as  Pauli spin operator-$1/2$.   $H_{i}$ is the random magnetic field that obeys the following bimodal distribution:%
\begin{equation}
\mathcal{P}(H_{i})=\frac{1}{2}\left[ \delta (H_{i}-H)+\delta (H_{i}+H)\right],  \tag{2}
\end{equation}%
in which $H\equiv \sqrt{\left\langle H_{i}^{2}\right\rangle _{c}}$ is the root mean square deviation of the magnetic field correspondent to the configurational average of the probability distribution $\mathcal{P}(H_{i})$.

The thermal average  of a general function involving spin operator components in a finite cluster $\mathcal{O}$(\{N\}) can be obtained by the   generalized relation of Callen and Suzuki~\cite{1}, i.e., %
\begin{equation}
\left\langle \mathcal{O}\mathbf{(\{N\})}\right\rangle =\left\langle \frac{%
\text{Tr}_{\{N\}}\left\{ \mathcal{O}\mathbf{(\{N\})e}^{-\beta H_{N}}\right\} 
}{\text{Tr}_{\{N\}}\left\{ \mathbf{e}^{-\beta H_{N}}\right\} }\right\rangle ,
\tag{3}
\end{equation}%
where the partial trace Tr$_{\{N\}}$ is taken over the set of $\mathbf{N}$ spin variables specified by a finite-system Hamiltonian $\mathcal{H}_{\mathbf{N}}$. \textbf{\  }$\left\langle \cdot \cdot \cdot \right\rangle $ indicates the canonical thermal average taken over the ensemble defined by the complete Hamiltonian (1).

The Callen-Suzuki identity for a finite cluster with two-spins was derived for the first time by Bob\'ak and Ja\u{s}\u{c}ur~\cite{bj} to study the criticality of the Ising model. It  has also been generalized for the description of the quantum spin-$1/2$ Heisenberg ferromagnet~\cite{10} and antiferromagnet~\cite{11}. De Sousa and Albuquerque~\cite{sa} (see also Refs. \cite{6, 7, 8}) have applied EFT-2 on the classical $n$-vector model. Latter, the EFT-2 approach was used to study the magnetic properties of the quantum spin-${1}$  Heisenberg ferromagnet~\cite{12}. More recently, this new EFT has been successfully used to treat second-order phase transitions of classical and quantum models~\cite{a,b,c,d,e}, and also to treat first-order transitions~\cite{f,g,h,i,j,l,m}.

Using a two-spin Hamiltonian for the finite system $\mathcal{H}_{2}$  in the Eq. (3) (see more details in Refs. \cite{6, 7,  8}), the magnetization per spin $\mathbf{m=}\left\langle \frac{1}{2}\left(S_{1}^{z}+S_{2}^{z}\right) \right\rangle $ is found. Applying the differential operator technique and EFT, an approximate expression for $\mathbf{m}$ is obtained for all values of $\mathbf{z}$. In particular, for the simple cubic lattice ($\mathbf{z=6}$) case, the average magnetization $\mathbf{m}$ is given by the following
expression:
\begin{equation}
m=\Lambda (m,T,H)=\sum\limits_{r=0}^{4}A_{2r+1}(T,H)m^{2r+1},  \tag{4}
\end{equation}%
where%
\begin{equation}
\Lambda (m,T,H)=\left[ \left( \alpha _{x}+m\beta _{x}\right) \cdot \left(
\alpha _{y}+m\beta _{y}\right) \right] ^{5}\left. G(x,y)\right\vert _{x,y=0},
\tag{5}
\end{equation}%
\begin{equation}
G(x,y)=\frac{1}{2}\left[ G_{+}^{c,q}(x,y)+G_{-}^{c,q}(x,y)\right] ,  \tag{6}
\end{equation}%
\begin{equation}
G_{\pm }^{q}(x,y)=\frac{\sinh (x+y\pm 2h)}{\cosh (x+y\pm 2h)+e^{2K}\cosh 
\sqrt{(x-y)^{2}+4K^{2}}}\hfill \text{ (quantum case),}  \tag{7}
\end{equation}%
\begin{equation}
G_{\pm }^{c}(x,y)=\frac{\sinh (x+y\pm 2h)}{\cosh (x+y\pm 2h)+\phi (K)\cosh
(x-y)}\text{\ (classical case)},  \tag{8}
\end{equation}%
\begin{equation}
\phi (K)=\frac{1-\mathcal{L}(3K)}{1+\mathcal{L}(3K)},  \tag{9}
\end{equation}%
\begin{equation}
A_{p}(T,H)=\frac{1}{p!}\left( \frac{\partial ^{p}\Lambda (m,T,H)}{\partial
m^{p}}\right) _{m=0},  \tag{10}
\end{equation}%
and%
\begin{equation}
\mathcal{L}(x)=\coth (x)-1/x\text{ (\textbf{Langevin function})}  \tag{11}
\end{equation}%
where $\alpha _{\mu }=\cosh (KD_{\mu })$, $\beta _{\mu }=\sinh (KD_{\mu })$ (%
$\mu =x,y$), $D_{\mu }=\frac{\partial }{\partial \mu }$ is the differential
operator, $K=J/k_{B}T$ and $h=H/k_{B}T$. The coefficients $A_{r}(T,H)$, Eq.
(10), are determined by applying the identity $%
e^{aD_{x}+bD_{y}}G(x,y)=G(x+a,y+b)$, and other corresponding expressions that are rather lengthy to be reproduced here.

\section{Results and Conclusions}
The EFT-2 was developed for the quantum spin-$1/2$ Heisenberg~\cite{10} and classical spin~\cite{11}(see also Refs. \cite{6,7, 8}) 
ferromagnet. Therefore, the expression from Eq. (4) has been obtained. This new method (EFT-2)  was also used to study the criticality 
of the quantum spin-$1$ anisotropic Heisenberg ferromagnet\cite{12}. It has been observed, from these works, that the critical 
temperature $k_{B}T_{c}/J$ increases with increasing spin ($S$) value, i.e., $k_{B}T_{c}/J\simeq 1.222$, $3.434$, and  $5.030$ 
for $S=1/2$, $1$, and $\infty $, respectively. These critical behavior for the dependence of $T_{c}$ with the value of the spin $S$, 
our results confirm the known results of series expansion~\cite{series}, where the values found are $k_{B}T_{c}/J\simeq 0.830$, $2.72$, and 
$4.329$, for $S=1/2$, $1$, and $\infty $, respectively. For a continuous phase transition, 
$m(T,H)$ decreases as the temperature increases and at $T=T_{c}(H)$ the order parameter is null (continuously). 
Then from Eq. (4) one can locate the second-order line through the condition%
\begin{equation}
A_{1}(T_{c},H)=1, \tag{12}
\end{equation}%
with $A_{3}(T_{c},H)>0$, and, additionally, the tricritical point can be located when%
\begin{equation}
A_{3}(T_{c},H)=0, \tag{13}
\end{equation}%
with $A_{5}(T_{c},H)<0$.
Depending on the range of the ratio $\delta =H/J$, we have second-order ($ 0<\delta <\delta _{t}$) and first-order ($\delta >\delta _{t}$) transitions, where ($\delta _{t},T_{t}$) is the tricritical point. One can note that it is not possible to calculate first-order transition lines in the basis of only the equation of state (4)  because in this case $m\neq 0$ at the transition point. To solve this problem one needs to compute the free energy for the ferromagnetic (F) and paramagnetic (P) phases. First-order transitions then correspond to locus on the phase diagram where free energies are equal. Assuming that the equation of state (4) is obtained by the minimization of a given free energy functional like $\Psi (m,T,H)$ (i.e., $\frac{\partial \Psi }{\partial m}=0$), we can express such relation as%
\begin{equation}
\Psi (m,T,H)=\lambda _{0}(T,H)+\frac{\lambda _{1}(T,H)}{2}\left[
1-\sum\limits_{r=0}^{4}\frac{A_{2r+1}(T,H)}{r+1}m^{2r}\right] m^{2}, 
\tag{14}
\end{equation}%
where $\lambda _{0}(T,H)$ and \ $\lambda _{1}(T,H)$ are arbitrary functions which turn out to be irrelevant when searching for the phase transition.  The Eq. (14) just represents qualitatively a Landau-like expansion, that can not be used to obtain the thermodynamic properties, only to study the phase diagram of spin system. This purpose for the free energy functional has been recently applied with success to study spin systems with frustration\cite{f,g,h,i,j, l, m}. In the present paper, we use it in the random field Heisenberg model to certify the potentiality of the methodology. It is known that this Landau expansion for $m$ is given by a finite serie and it is possible to show that $\lambda _{1}(T,H)>0.$ Thus, we assume that this parameter $\lambda _{1}(T,H)$ is also positive in Eq. (14). Near the criticality (i.e., $T\simeq T_{c}$, $m\simeq 0$) we have, from the equation of state (4), the behavior of the magnetization  given by $m\simeq \sqrt{\frac{1-A_{1}(T,H)}{A_{3}(T,H)}\text{ }}$ (classical critical exponent, $\beta =1/2$) and, consequently from Eq. (14) $\frac{\partial ^{2}\Psi }{\partial m^{2}}\simeq -2\left[ 1-A_{1}(T,H)\right] >0$ that corresponds to a \textit{minimum} point (stability limit). We note that $A_{3}(T,H)>0$ and $A_{1}(T,H)>1$ for all $%
H<H_{t}$ (tricritical field) and $T<T_{c}$.
From Eq. (14), we obtain the separation point  of the two phases F ($m\neq 0$) and P ($m=0$), i.e., $\Psi _{F}(m,T,H)=\Psi _{P}(0,T,H)$
\begin{equation}
\sum\limits_{r=0}^{4}\frac{A_{2r+1}(T,H)}{r+1}m^{2r}=1 \tag{15}
\end{equation}
In Refs. \cite{6,8}, the Eqs. (12) and (13) have been used to obtain the critical frontier which separates the F phase from the P phase and the tricritical point (TCP) for the \textbf{classical} and quantum cases. Simultaneously solving Eqs. (4) and (15) we obtain the second-order line when $m=0$ and first-order line when $m\neq 0$. The corresponding phase diagram in the $T-H$ plane is depicted in Figure~1 for the classical and quantum spins. As a first observation, we  note that the nature of variations of $T_{c}$ versus $H$ reveal a common basic behavior -    the transition temperature  decreases when $H/J$ increases, reaching the zero temperature limit at same value of $H_{c}/J$ (i.e., $H_{c}/J=3.23$). We have also observed  that $T_{c}(H)$ for the classical case is larger than the quantum case, what is accepted physically. 

In conclusion, we observe that EFT formalism allows us to study the random field Heisenberg (classical and quantum) model with correlation and phase diagram in the $T-H$ plane. The results by using the functional for free energy are satisfactory to calculate the first-order line with qualitative and, to a certain extent, quantitative confidence. We can also extend the presented methodology  to study the magnetic properties\cite{thermal1, thermal2}.

\section*{ACKNOWLEDGMENTS}

J. R. de Sousa and A. S. de Arruda are partially supported by CNPq and Fapemat respectively  Brazilian agency.

\newpage
\begin{figure}[!h]
\centering
\includegraphics[scale=0.8]{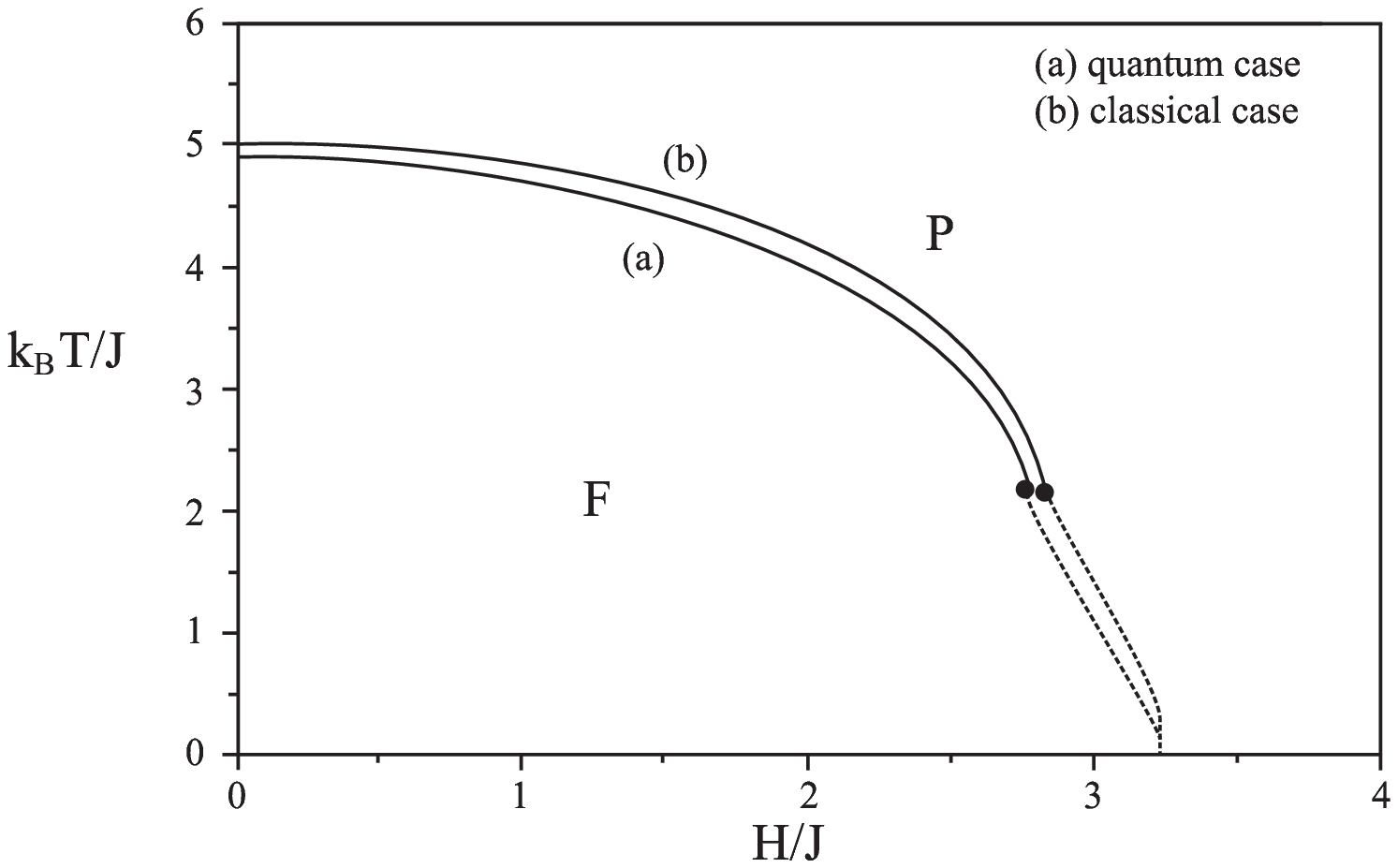}
\caption{Phase diagram in the $T-H$ plane of the random field
Heisenberg model on a simple cubic lattice for quantum (a) and classical (b)
spin cases. The solid and dashed lines correspond to the second- and first-
order phase transition respectively. The tricritical point is marked by a
back point.}
\label{figura}
\end{figure}

\end{document}